\documentclass{elsart}

\usepackage{epsfig}

\usepackage{amssymb}

\begin{document}

\begin{frontmatter}

\title{Neutron and proton energy spectra from the non-mesonic weak decays of
 $^{5}_{\it{\Lambda}}$He and $^{12}_{\it{\Lambda}}$C}
\author[titech]{S. Okada\thanksref{NOWriken}},
\ead{sokada@riken.jp}
\author[osaka]{S. Ajimura},
\author[kek]{K. Aoki},
\author[gsi]{A. Banu},
\author[snu]{H. C. Bhang},
\author[kek]{T. Fukuda\thanksref{NOWoecu}},
\author[tohoku]{O. Hashimoto},
\author[snu]{J. I. Hwang},
\author[tohoku]{S. Kameoka},
\author[snu]{B. H. Kang},
\author[snu]{E. H. Kim},
\author[snu]{J. H. Kim\thanksref{NOWjungho}},
\author[snu]{M. J. Kim},
\author[ut]{T. Maruta},
\author[tohoku]{Y. Miura},
\author[osaka]{Y. Miyake},
\author[kek]{T. Nagae},
\author[ut]{M. Nakamura}, 
\author[tohoku]{S. N. Nakamura},
\author[kek]{H. Noumi},
\author[tohoku]{Y. Okayasu},
\author[kek]{H. Outa\thanksref{NOWriken}},
\author[kriss]{H. Park},
\author[kek]{P. K. Saha\thanksref{NOWjaeri}},
\author[kek]{Y. Sato}, 
\author[kek]{M. Sekimoto},
\author[tohoku]{T. Takahashi\thanksref{NOWkek}},
\author[tohoku]{H. Tamura},
\author[riken]{K. Tanida},
\author[kek]{A. Toyoda}, 
\author[tohoku]{K. Tsukada},
\author[tohoku]{T. Watanabe},
\author[snu]{H. J. Yim}

\address[titech]{Department of Physics, Tokyo Institute of Technology,
 Ookayama 152-8551, Japan}
\address[osaka]{Department of Physics, Osaka University,
 Toyonaka 560-0043, Japan}
\address[kek]{High Energy Accelerator Research Organization (KEK),
 Tsukuba 305-0801, Japan}
\address[gsi]{Gesellschaft f$\ddot{\mbox{u}}$r Schwerionenforschung mbH (GSI),
 Darmstadt 64291, Germany}
\address[snu]{Department of Physics, Seoul National University,
 Seoul 151-742, Korea}
\address[tohoku]{Department of Physics, Tohoku University,
 Sendai 980-8578, Japan}
\address[ut]{Department of Physics, University of Tokyo,
 Hongo 113-0033, Japan}
\address[kriss]{Korea Research Institute of Standards and Science (KRISS),
 Daejeon 305-600, Korea}
\address[riken]{RIKEN Wako Institute, RIKEN,
 Wako 351-0198, Japan}

\thanks[NOWriken]{Present address: RIKEN Wako Institute, RIKEN,
 Wako 351-0198, Japan}
\thanks[NOWoecu]{Present address: Laboratory of Physics,
 Osaka Electro-Communication University,
 Neyagawa 572-8530, Japan}
\thanks[NOWjungho]{Present address: Department of Physics,
 Chung-Ang University, Seoul 143-747, Korea.}
\thanks[NOWjaeri]{Present address:
 Japan Atomic Energy Research Institute, Tokai 319-1195, Japan}
\thanks[NOWkek]{Present address:
 High Energy Accelerator Research Organization (KEK),
 Tsukuba 305-0801, Japan}

\begin{abstract}

We have simultaneously
measured the energy spectra of neutrons and protons
emitted in the non-mesonic weak decays
of $^{5}_{\it{\Lambda}}$He and $^{12}_{\it{\Lambda}}$C hypernuclei
produced via the ($\pi^+$,$K^+$) reaction
with much higher statistics over those of previous experiments.
The neutron-to-proton yield ratios
for both hypernuclei
at a high energy threshold (60 MeV) were approximately equal to two,
which suggests that the ratio of the neutron- and
proton-induced decay channels,
$\Gamma_n$($\it{\Lambda}n \to nn$)/$\Gamma_p$($\it{\Lambda}p \to np$),
is about 0.5.
In the neutron energy spectra,
we found that the yield of the low-energy component is
unexpectedly large, even for $^{5}_{\it{\Lambda}}$He.
 
\end{abstract}

\begin{keyword}
 $\it{\Lambda}$ hypernuclei \sep
 non-mesonic weak decay \sep
 nucleon energy spectrum
 \PACS 
 21.80.+a \sep
 13.30.Eg \sep
 13.75.Ev
\end{keyword}
\end{frontmatter}

\section{Introduction}

In free space, a $\it{\Lambda}$ particle
decays dominantly associated with a pion
in the final state as $\it{\Lambda}$ $\to N \pi$.
In the case of $\it{\Lambda}$ bound in a nucleus,
a $\it{\Lambda}$ hypernucleus,
$\it{\Lambda}$ is not only possible to decay as in the free space 
(mesonic weak decay),
but is also able to stimulate a nucleon in the nucleus
as $\it{\Lambda}$``$N$''$\to n N$ (non-mesonic weak decay, NMWD).
The NMWD gives a unique opportunity
to study the weak interaction between baryons,
because this strangeness non-conserving process is purely attributed to
the weak interaction.

In the NMWD of a $\it{\Lambda}$ hypernucleus,
there are two decay channels,
$\it{\Lambda}p$ $\to n p$ ($\Gamma_p$) and
$\it{\Lambda}n$ $\to n n$ ($\Gamma_n$).
The ratio of those decay widths,
$\Gamma_n$/$\Gamma_p$,
is an important observable
used to study the isospin structure of the NMWD mechanism.
For the past 40 years, there has been a longstanding puzzle
that the experimental $\Gamma_n$/$\Gamma_p$ ratio
disagrees with that of theoretical calculations
based on the most natural and simplest model,
the One-Pion Exchange model (OPE).
In this model, the $\it{\Lambda} N \to nN$ reaction is expressed
as a pion absorption process 
after the $\it{\Lambda} \to N \pi$ decay inside the nucleus.
Since the OPE process is tensor-dominant
and the tenser transition of the initial $\it{\Lambda} N$ pair
in the $s$-state requires the final $n N$ pair to have isospin zero,
the $\Gamma_n/\Gamma_p$ ratio in OPE process becomes close to 0.
However, previous experimental results have indicated
a large $\Gamma_n/\Gamma_p$ ratio ($\sim$1) \cite{Szy91,Noumi95}.

This large discrepancy between the OPE-model predictions and
the experimental results has stimulated many theoretical studies:
the heavy meson exchange model \cite{Dub86,Par97,Par02},
the Direct Quark model \cite{Che83,Hed86,Ino96,Ino98,Sas00}
and the two-nucleon (2$N$) induced model
($\it{\Lambda} N N \rightarrow n N N$) \cite{Alb91}.
After K. Sasaki $\textit{et~al.}$
pointed out an error about the sign
of the Kaon exchange amplitudes in 2000 \cite{Sas00},
those theoretical values of the $\Gamma_n/\Gamma_p$ ratio
have increased to the level of 0.4$\sim$0.7 \cite{Alb02}.

On the other hand, the experimental data still have large errors
($\Gamma_n/\Gamma_p$ = 0.93 $\pm$ 0.55
for $^{5}_{\it{\Lambda}}$He \cite{Szy91}),
and it is hard to draw a definite conclusion
on the $\Gamma_n/\Gamma_p$ ratio.
So far, several nucleon energy spectra from hypernuclear decay
have been reported.
Most of the experiments measured only the proton energy spectra,
because of the difficulty in detecting neutrons.
The $\Gamma_n/\Gamma_p$ ratio
has been estimated through a comparison of the measured proton
spectrum with that of the intra-nuclear cascade calculation
by changing the $\Gamma_n/\Gamma_p$ ratio \cite{Sat03}.
In this method, there are large
uncertainties due to the indirect evaluation of neutrons.

Recently, the neutron energy spectra from $^{12}_{\it{\Lambda}}$C and 
$^{89}_{\it{\Lambda}}$Y were measured with high statistics \cite{Kim03}.
For $^{12}_{\it{\Lambda}}$C, the $\Gamma_n/\Gamma_p$ ratio
was extracted from the neuron-to-proton yield ratio, $N_n/N_p$,
estimated by the neutron energy spectrum
combined with a proton energy spectrum measured 
in another experiment \cite{Sat03}.
The energy loss of protons in the target, however,
was not able to be corrected for.
Since there were ambiguities coming from different detection energy
thresholds in the two measurements, the value might be affected
by a large systematic error.
We therefore simultaneously measured the neutron and proton spectra
from the decay of $\it{\Lambda}$ hypernuclei
with a proton energy loss correction, and
with much higher statistics over those of previous experiments.

When we compare the ``measured'' $\Gamma_n/\Gamma_p$ ratio
with that obtained in theoretical calculations,
the most serious technical problem was a treatment
of the re-scattering effect in the residual nucleus,
the so-called Final State Interaction (FSI).
The nucleon energy spectrum
emitted from the two-body NMWD process, $\it{\Lambda} N \to nN$,
must originally have a broad peak
at half of the Q-value ($\sim$ 76 MeV).
The FSI effect distorts the nucleon spectra
enhancing the low energy region,
in which the strength of the effect depends on
the mass number of the hypernuclei.
In order to minimize the FSI effect,
we selected a light $s$-shell hypernucleus, $^{5}_{\it{\Lambda}}$He.
In the $s$-shell hypernucleus,
initial relative $\it{\Lambda} N$ states must be $S$ states,
whereas in a $p$-shell hypernucleus it is possible to be $P$ states.
According to Ref.\ \cite{Ben92},
the transitions coming from the $P$ states were predicted to be small,
though we need experimental confirmation for that.
In order to investigate the $p$-wave effect,
we also performed the same experiment
for a typical light $p$-shell hypernucleus, $^{12}_{\it{\Lambda}}$C.
In this paper, we show those energy spectra for both hypernuclei and
present the neuron-to-proton yield ratios, $N_n/N_p$,
with a high energy threshold (60 MeV)
in order to get rid of the contribution from the FSI effect.

On the other hand,
the nucleon yields in the low energy region should be sensitive to
the strength of the FSI effect.
In the proton measurement there is a cutoff energy,
typically at 30 $\sim$ 40 MeV,
due to the energy loss inside the target and the detector,
whereas in the neutron measurement
we have experienced sensitivity in the whole energy region.
In this paper, we also show the mass number dependence
of the ``neutron'' energy spectra for A = 5 and 12,
with that for A = 89 \cite{Kim03},
which provides valuable information to investigate
the FSI effect. 
The possible existence of a multi-nucleon induced process
in which the Q-value can be distributed over three or more nucleons
in the final state has been discussed theoretically
(such as 2$N$-induced process),
though there has been no experimental evidence so far;
the nucleons emitted from this process
should also distribute in the low energy region.
Thus, the whole shape of the neutron energy spectrum also brings
useful knowledge on the multi-nucleon induced NMWD mechanism.

\section{Experimental method}

The present experiments (KEK-PS E462/E508) were carried out
at the 12-GeV proton synchrotron (PS)
in the High Energy Accelerator Research Organization (KEK).
Hypernuclei, $^{5}_{\it{\Lambda}}$He and $^{12}_{\it{\Lambda}}$C,
were produced via the ($\pi^+$,$K^+$) reaction at 1.05 GeV/$c$
on $^6$Li and $^{12}$C targets, respectively.
Since the ground state (g.s.) of $^{6}_{\it{\Lambda}}$Li
is above the threshold of $^{5}_{\it{\Lambda}}$He $+\ p$,
the g.s. promptly decays into
$^{5}_{\it{\Lambda}}$He emitting a low-energy proton.
The $^{6}$Li ($\pi^+$,$K^+$) $^{6}_{\it{\Lambda}}$Li reaction
was therefore employed to produce $^{5}_{\it{\Lambda}}$He.
The hypernuclear mass spectra were calculated by
reconstructing the momenta of incoming $\pi^+$ and outgoing $K^+$
using a beam-line spectrometer composed of the QQDQQ system
and the superconducting kaon spectrometer (SKS) \cite{Fuk95},
respectively.

Particles emitted from the decays of $\it{\Lambda}$ hypernuclei
were detected by the decay-particle detection system,
as shown in Fig.\ \ref{fig:e462setup}.
It was composed of plastic scintillation counters (T1, T2, T3 and T4)
and a multi-wire drift chamber (PDC).
The neutral particles, neutrons and gammas, were identified
by means of the time-of-flight (TOF) technique
between T1 and T4;
in addition, T3 was used as a veto counter
in order to reject charged particles.
The charged particles, protons and charged pions, were identified
by utilizing $dE/dx$ on T2,
total energy deposit on sequentially fired counters (T2, T3 and T4),
and the TOF between T2 and T3.
Their trajectories were measured by PDC.

\begin{center}
 \makeatletter
 \def\@captype{figure}
 \makeatother
 \epsfig{file=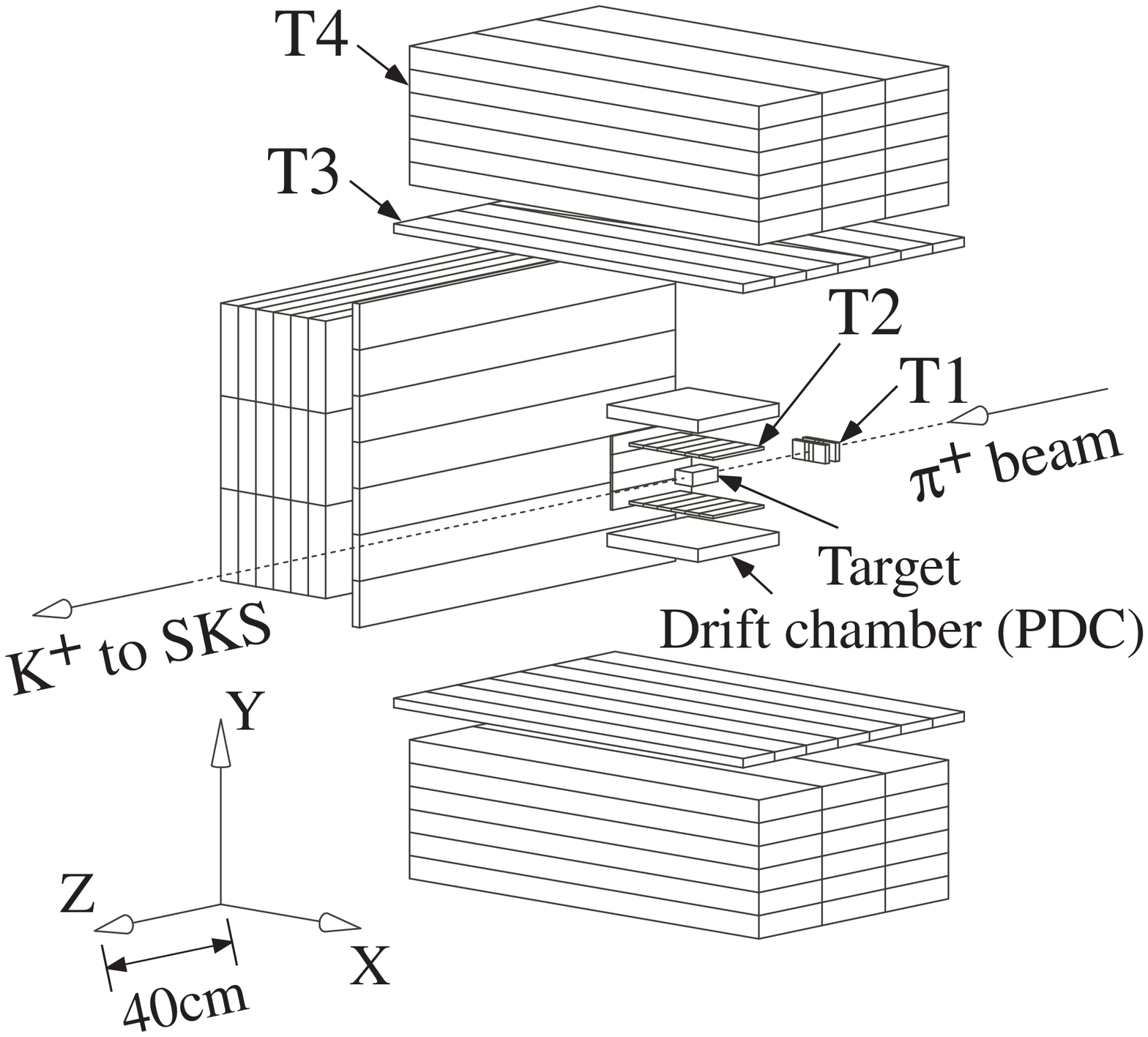,width=6.5cm}
 \caption{Schematic view
 of the decay-particle detection system.}
 \label{fig:e462setup}
\end{center}

\section{Analysis}

Figure \ref{fig:ex} shows
the inclusive excitation-energy spectra
of $^{6}_{\it{\Lambda}}$Li and $^{12}_{\it{\Lambda}}$C.
The g.s. peak is clearly seen
in our excitation spectra for both hypernuclei.
We selected events from the hypernuclear production
by gating the g.s. peak, as shown in the figure.

The yields of $^{5}_{\it{\Lambda}}$He and $^{12}_{\it{\Lambda}}$C
are, respectively, about 4.6 $\times$ 10$^4$ and 6.2 $\times$ 10$^4$,
which were one order-of-magnitude higher
than those of previous experiments.
There is a constant background below the g.s. peak;
the background for $^{5}_{\it{\Lambda}}$He and $^{12}_{\it{\Lambda}}$C
within the g.s. gate
were estimated to be as small as 8\% and 3\%, respectively.
We estimated the contribution
from this constant background in the nucleon spectra
and subtracted it from the g.s. gated nucleon spectra.

\begin{center}
 \makeatletter
 \def\@captype{figure}
 \makeatother
 \epsfig{file=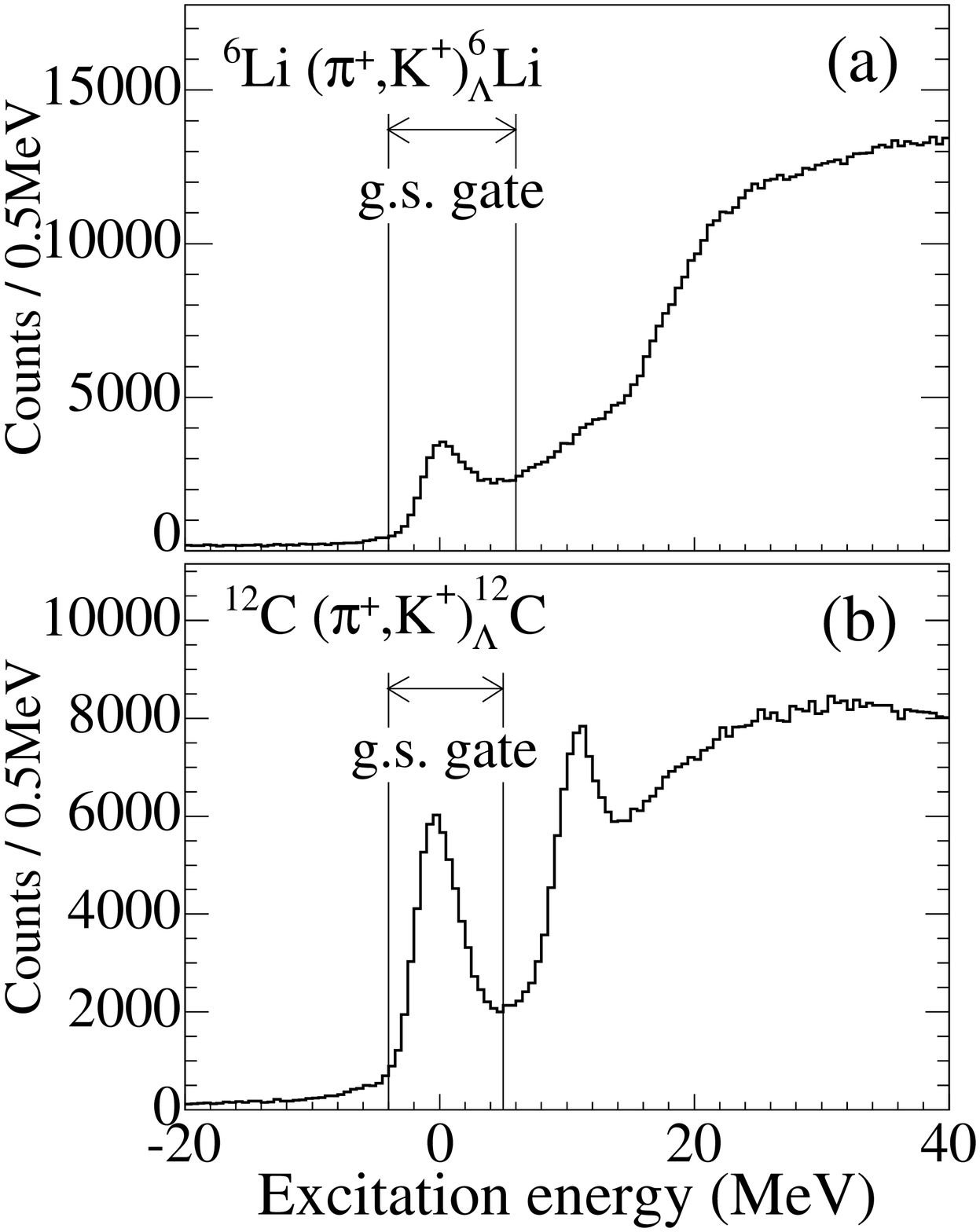,width=6.5cm}
 \caption{Inclusive excitation-energy spectra of
 (a) $^{6}_{\it{\Lambda}}$Li and (b) $^{12}_{\it{\Lambda}}$C.}
 \label{fig:ex}
\end{center}

In the neutral decay particle analysis,
the TOF between the reaction vertex and the T4 counter, ${\it TOF}_n$,
is expressed as
${\it TOF}_n = T_{T4} - T_{T1} - {\it TOF}_{b} - \Delta\tau $,
where $T_{T4}$ and $T_{T1}$ are the timings
of T4 and T1 (time zero), respectively;
${\it TOF}_{b}$ means the flight time of the beam pion;
$\Delta\tau$ corresponds to
the time delay due to the hypernuclear lifetime.
Since $\Delta\tau$ cannot be determined for each event individually,
we replaced $\Delta\tau$ with the mean lifetime
of the hypernuclei to avoid a systematic energy
shift in the neutron energy spectra.
Here, we used the lifetimes of
$\tau = 278^{+11}_{-10}$ ps for $^{5}_{\it{\Lambda}}$He and
$\tau = 212^{+7}_{-6}$ ps for $^{12}_{\it{\Lambda}}$C,
improved through the lifetime analysis of the present experiment \cite{Kam03}.

The inverse of the neutron velocity, 1/$\beta$, was calculated
as 1/$\beta$ = ${\it TOF}_n$ / $L \times c$,
where $L$ is the flight length and $c$ is the light velocity.
The top of Fig.\ \ref{fig:pid} shows
the measured 1/$\beta$ spectra 
with a light-output threshold of 2 MeVee (MeV electron equivalent)
for $^{12}_{\it{\Lambda}}$C.
The energy scale is displayed
on the upper horizontal axis.
The gate region of the neutron was defined as 1.97 $< 1/\beta <$ 9.73,
which corresponds to 5 $<E_n<$ 150 MeV of neutron energy.
The spectrum shows good $\gamma$/neutron separation.
The extremely low background level
of the $1/\beta <$ 0 region in the spectrum
indicates that the accidental background is negligibly small.
The neutron energy resolutions for
$^{5}_{\it{\Lambda}}$He and $^{12}_{\it{\Lambda}}$C
were estimated from the width of the $\gamma$-ray peak.
They were, respectively, about 10 MeV and 7 MeV
in full width at half maximum (FWHM) at $E_n=$75 MeV.

The bottom of Fig.\ \ref{fig:pid} shows
the particle identification (PID) of the charged particles
emitted from the decay of $^{5}_{\it{\Lambda}}$He.
This plot was made by the PID function
derived from the $dE/dx$, the total energy and the particle velocity.
The charged particles
(charged pions, protons and deuterons) were separated well.
It is noteworthy that
deuterons were separated from the protons
for the first time in such counter experiments for the NMWD.
The proton peak was fitted by a Gaussian function,
and the proton gate was set at the $\pm$ 2 $\sigma$ region
of the Gaussian, as shown in the figure.
The loss of protons (4.6\%) by this tight cut was taken into account
in the analysis efficiency.
The proton kinetic energy was determined by the TOF between T2 and T3,
which must be affected by the energy losses
inside the target and in the T2 counter.
Those energy losses were estimated event-by-event
from the path length inside the target and the T2 counter,
which was calculated from the reaction vertex and the trajectory.

\begin{center}
 \makeatletter
 \def\@captype{figure}
 \makeatother
 \epsfig{file=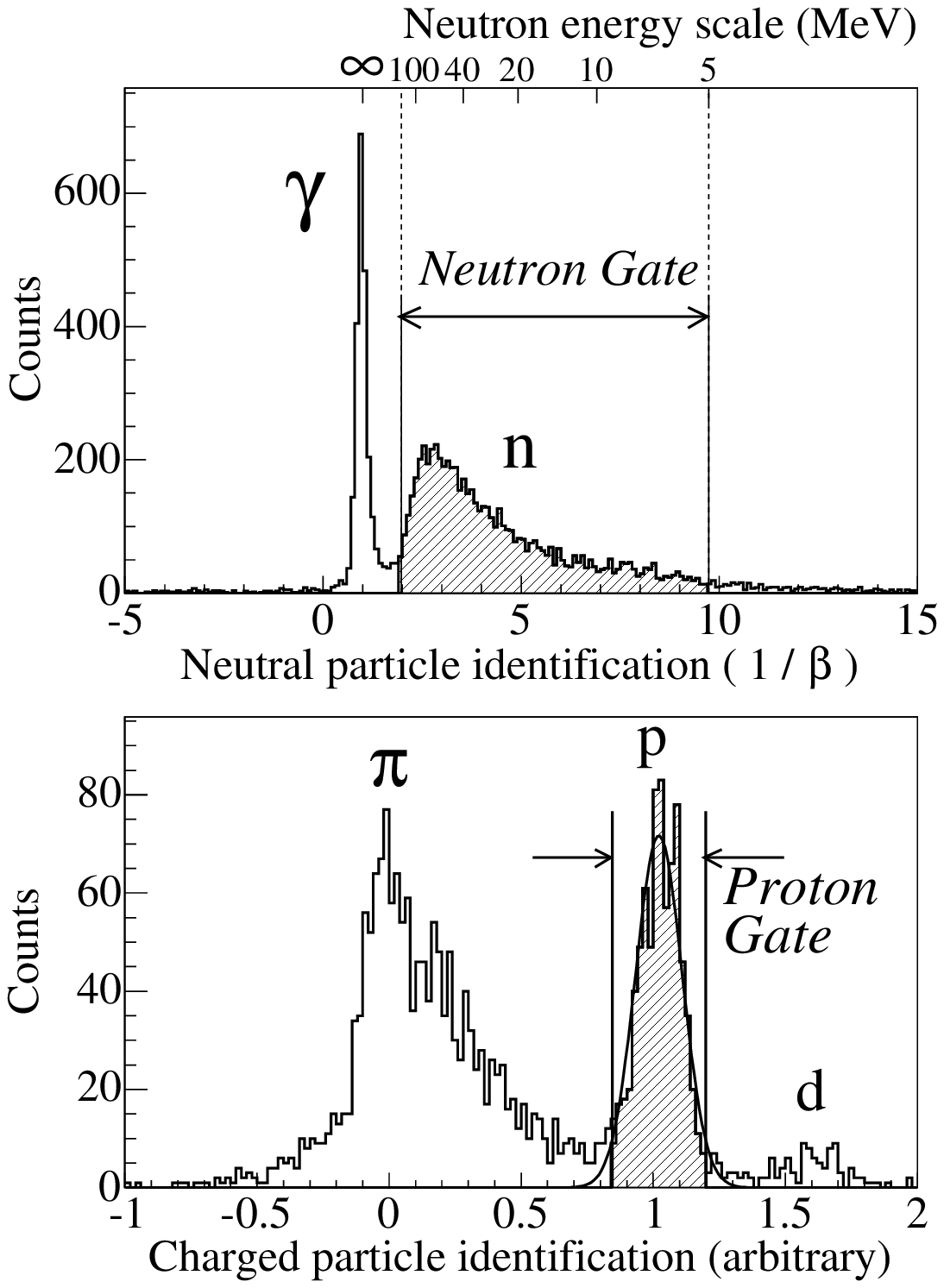,width=6.5cm}
 \caption{Particle identifications for neutral (top) and
 charged (bottom) decay particles.
 The top shows the 1/$\beta$ spectra of neutral particles
 emitted from the decay of $^{12}_{\it{\Lambda}}$C
 with 2 MeVee threshold.
 The bottom shows the PID function of charged particles
 emitted from the decay of $^{5}_{\it{\Lambda}}$He.}
 \label{fig:pid}
\end{center}

The numbers of the neutron and proton per NMWD at energy $E$,
$N_{n}(E)$ and $N_{p}(E)$, are expressed as
\begin{eqnarray}
 \label{eq:Nneu}
  N_{n,p}(E) ~=~ \frac{Y_{n,p}(E)}{Y_{HY} \cdot \Omega_{n,p} \cdot b_{nm}
  \cdot \varepsilon_{n,p}(E)} ~~,
\end{eqnarray}
where $Y_{HY}$ is the number of g.s. formations
in the inclusive spectrum;
$Y_{n}(E)$ and $Y_{p}(E)$ are the numbers of
detected neutrons and protons from the decay.

The acceptances for the neutron and proton detection system,
$\Omega_n$ and $\Omega_p$, were estimated respectively
to be 26.6 $\pm$ 0.8\% and 10.0 $\pm$ 0.3\% by a Monte-Carlo simulation.
Only for the proton analysis,
we selected the central parts of the segmented T2 and T3 counters
of the top and bottom coincidence arms
in order to suppress ambiguity of the acceptance estimation
due to escape from the side edge of the T4 counter.
The difference between $\Omega_n$ and $\Omega_p$ is attributed to
this selection.

The non-mesonic decay branching ratios of $^{5}_{\it{\Lambda}}$He
and $^{12}_{\it{\Lambda}}$C, $b_{nm}$, were evaluated as 
0.429 $\pm$ 0.012(stat) $\pm$ 0.005(sys) and
0.768 $\pm$ 0.005(stat) $\pm$ 0.012(sys),
respectively, by $\pi^-$ and $\pi^0$ branching-ratio analyses
of the present experiment \cite{Kam03,Oka03}.

The energy-dependent detection efficiency for the neutron,
$\varepsilon_n(E)$, was estimated
by the Monte-Carlo simulation code DEMONS \cite{Byrd92},
which is based on CECIL \cite{Cec79}
and is applicable to multi-element neutron detectors.
We compared the calculated efficiencies with those of
various existing experimental data \cite{Kim03},
and found that the systematic error does not exceed 6\%
for integrated yields above 10, 20, 30 and 40 MeV.
According to a Monte-Carlo simulation for the proton,
there is no experimental sensitivity to a proton below about 30 MeV,
due to the energy loss,
and the detection efficiency was gradually reduced
from 50 MeV to 30 MeV.
The energy-dependent detection efficiency for the proton,
$\varepsilon_p(E)$, was calculated using the simulation.

There is a non-negligible nucleon background
due to the pion absorption process
in which $\pi^-$'s from the mesonic decay of $\it{\Lambda}$ hypernucleus
are absorbed by the materials around the target.
The background was estimated
by assuming that the shape of the nucleon spectra
from this $\pi^-$ absorption process is the same as
that from the $\pi^-$ decay of $\it{\Lambda}$
($\it{\Lambda}$ $\to \pi^- p $) formed
via the quasi-free formation process.
Here, we regarded
the $\pi^-$ branching ratio of the quasi-free $\it{\Lambda}$ decay
as that of $\it{\Lambda}$ in the free space,
and estimated the contribution using the $\pi^-$ branching ratios,
0.359 $\pm$ 0.009 for $^{5}_{\it{\Lambda}}$He \cite{Kam03}
and 0.099 $\pm$ 0.011(stat) $\pm$ 0.004(sys) for
$^{12}_{\it{\Lambda}}$C \cite{Sat03}.

\section{Results and Discussion}

The top and middle of Fig.\ \ref{fig:npene} show
the neutron/proton energy spectra from the decay of
$^{5}_{\it{\Lambda}}$He and $^{12}_{\it{\Lambda}}$C, respectively.
They were normalized per NMWD.
It seems that the neutron spectra for both hypernuclei
have a similar shape to those of protons
above the proton energy threshold of 30 MeV,
and the neutron yields were about twice higher than those of protons.
Thanks to the event-by-event energy-loss correction for proton,
we can directly compare the neutron energy spectra
with those of the protons at the same energy threshold.

\begin{center}
 \makeatletter
 \def\@captype{figure}
 \makeatother
 \epsfig{file=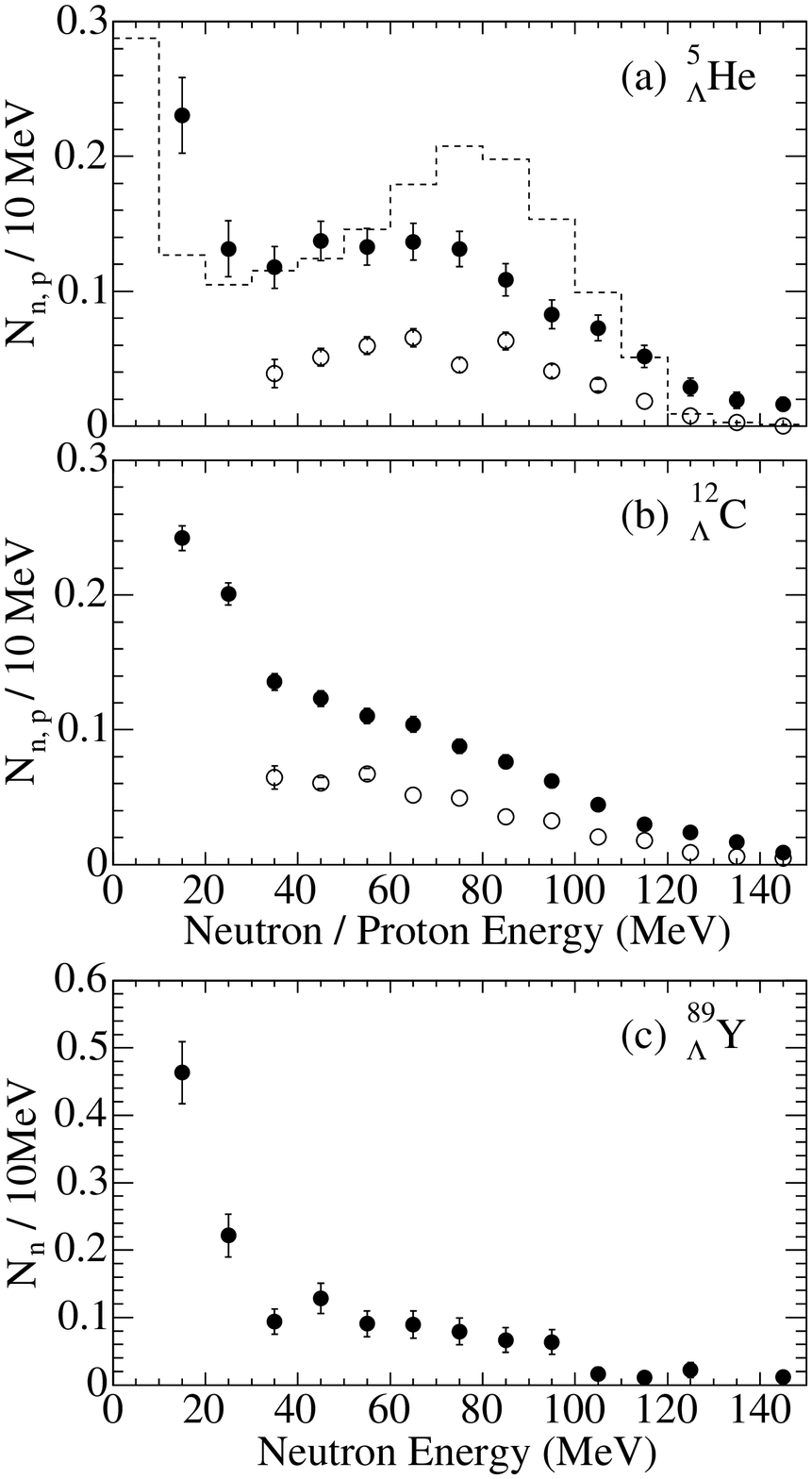,width=6.5cm}
 \caption{Neutron (filled circle) and proton (open circle)
 energy spectra per NMWD of 
 (a) $^{5}_{\it{\Lambda}}$He and (b) $^{12}_{\it{\Lambda}}$C,
 with the neutron spectrum of (c) $^{89}_{\it{\Lambda}}$Y
 obtained in the previous experiment \cite{Kim03}.
 The errors are statistical.
 The dashed histogram in the top figure shows the neutron spectrum
 per NMWD of $^{5}_{\it{\Lambda}}$He,
 calculated by Garbarino $\textit{et~al.}$,
 in which the FSI effect and the 2$N$-induced process
 were taken into account \cite{Gar03}.}
 \label{fig:npene}
\end{center}

It should be noted that the partial decay rate of
the $^{5}_{\it{\Lambda}}$He $\to n + \alpha$ process,
which emits a monochromatic high-energy neutron ($\sim$ 135 MeV),
is not negligible.
The decay rate was reported
as 0.049 $\pm$ 0.01 $\Gamma_{\pi^-}$  \cite{Cor70},
where $\Gamma_{\pi^-}$ is the $\pi^-$ decay width
of $^{5}_{\it{\Lambda}}$He.
Though it is hard to observe the corresponding peak
due to the limited resolution, $\sim 25$ MeV (FWHM)
at such a high-energy region,
one can see an appreciable neutron yield
at the high-energy region (120 $\sim$ 150 MeV)
in the neutron spectra of $^{5}_{\it{\Lambda}}$He,
which can be considered to be the events from this process.

The FSI process and the possible multi-nucleon induced process
(such as the 2$N$-induced process)
tend to enhance the low-energy region in the nucleon energy spectra.
In order to reduce the effects from those processes,
we set the energy threshold to be as high as 60 MeV,
which is near to half of the Q-value ($\sim$ 76 MeV)
of the $\it{\Lambda} N \to nN$ decay process.
The ratios of the yields between neutrons and protons, $N_n/N_p$,
for $^{5}_{\it{\Lambda}}$He and $^{12}_{\it{\Lambda}}$C
above 60 MeV, were obtained respectively as follows:
\begin{eqnarray}
 N_{n}/N_{p}~(^{5}_{\it{\Lambda}}\mbox{He}) &=&
  2.17 \pm 0.15(\mbox{stat}) \pm 0.16(\mbox{sys})
  ~~(60<E<110\mbox{MeV}), \\
 N_{n}/N_{p}~(^{12}_{\it{\Lambda}}\mbox{C}) &=&
  2.00 \pm 0.09(\mbox{stat}) \pm 0.14(\mbox{sys}) ~~(E>60\mbox{MeV}),
\end{eqnarray}
where we set the upper energy limit of 110 MeV
only for $^{5}_{\it{\Lambda}}$He
in order to avoid the contamination
from the $^{5}_{\it{\Lambda}}$He $\to n + \alpha$ process.
Systematic errors are predominantly coming
from an ambiguity of the absolute neutron detection efficiency
($\sim$6\%).
Due to the similar spectrum shape for neutrons and protons,
the $N_n/N_p$ ratios of both hypernuclei
are quite stable for a change of the energy thresholds.
Recently, an independent experimental result of the $N_n/N_p$
on $^{12}_{\it{\Lambda}}$C was reported
as $1.73 \pm 0.22$ \cite{Kim03},
which agrees with our result within the errors.
If the FSI effect and the contribution
of the 2$N$-induced process can be neglected,
the relation between the $\Gamma_n/\Gamma_p$ and 
the $N_{n}/N_{p}$ for $\it{\Lambda} N \to nN$ decay process
is approximately expressed as
\begin{eqnarray}
 N_{n}/N_{p} \cong  2 \times ( \Gamma_n/\Gamma_p ) +  1.
  \label{RelationNnNpNnnNnp}
\end{eqnarray}
When we applied the above relation to the obtained $N_{n}/N_{p}$,
the $\Gamma_n/\Gamma_p$ ratios for both hypernuclei were
obtained as about 0.5$\sim$0.6.
The result excludes earlier experimental results that
the $\Gamma_n/\Gamma_p$ is close to unity \cite{Szy91,Noumi95}.
It also rules out a theoretical calculation based on the OPE model,
which predicts that the value should be as small as 0.1.
Our result supports recent calculations based on
short-range interactions, such as the heavy-meson exchange
and the direct quark exchange.
Moreover, we could not observe significant difference
of the $\Gamma_n/\Gamma_p$ ratio
between the $s$-shell and $p$-shell hypernuclei.
It is suggested small contribution from the $p$-wave in this process.

The bottom of Fig.\ \ref{fig:npene}
shows the neutron spectrum per NMWD of $^{89}_{\it{\Lambda}}$Y
obtained in a previous experiment \cite{Kim03}.
With this spectrum,
the mass-number dependence of the neutron energy spectra
for A = 5, 12 and 89 is shown.
The integrated numbers of neutrons per NMWD
of $^{5}_{\it{\Lambda}}$He, $^{12}_{\it{\Lambda}}$C
and $^{89}_{\it{\Lambda}}$Y are listed in Table \ref{table:NumNeuNMWD}
for energy thresholds of 10, 20, 30 and 40 MeV.
The high-energy region is suppressed with the increase of mass number
while enhancing the low-energy component.
This tendency can be naturally interpreted as being the FSI effect.

\begin{table}[htbp]
 \caption{Total number of neutrons per NMWD of 
 $^5_{\it{\Lambda}}$He, $^{12}_{\it{\Lambda}}$C and $^{89}_{\it{\Lambda}}$Y
 \cite{Kim03} with energy thresholds of 10, 20, 30 and 40 MeV.}
 \label{table:NumNeuNMWD}
 \begin{center}
  \begin{tabular}{c|cc|c}
   \hline \hline
   & \multicolumn{3}{c}{Total number of neutrons per non-mesonic decay} \\
   \hline 
   & $^5_{\it{\Lambda}}$He & $^{12}_{\it{\Lambda}}$C & $^{89}_{\it{\Lambda}}$Y \cite{Kim03}\\
   \hline
   $E_n >$ 10 MeV & 1.398$\pm$0.052$\pm$0.096 & 1.266$\pm$0.020$\pm$0.090 & 1.36$\pm$0.08$\pm$0.09  \\
   $E_n >$ 20 MeV & 1.168$\pm$0.044$\pm$0.080 & 1.024$\pm$0.018$\pm$0.072 & 0.89$\pm$0.06$\pm$0.06  \\
   $E_n >$ 30 MeV & 1.036$\pm$0.039$\pm$0.071 & 0.823$\pm$0.016$\pm$0.058 & 0.67$\pm$0.06$\pm$0.04  \\
   $E_n >$ 40 MeV & 0.918$\pm$0.036$\pm$0.063 & 0.687$\pm$0.015$\pm$0.048 & 0.58$\pm$0.05$\pm$0.04  \\
   \hline
   \hline
  \end{tabular}
 \end{center}
\end{table}

If the 1$N$-induced process ($\it{\Lambda} N \rightarrow n N$)
dominates the NMWD and the FSI effect is negligible,
the neutron energy spectrum should have a broadened peak,
due to the Fermi motion, at about one half of the Q-value
of the process ($\sim$75 MeV).
We simply expected that this would be the case for $^{5}_{\it{\Lambda}}$He
due to its small mass number.
Garbarino $\textit{et~al.}$ reported
a calculated nucleon energy spectrum for $^{5}_{\it{\Lambda}}$He
with the FSI effect and assuming that $\Gamma_n/\Gamma_p = 0.46$ and
$\Gamma_{2N}/\Gamma_{1N} = 0.20$,
where $\Gamma_{1N}$ and $\Gamma_{2N}$ denote the decay widths
of the 1$N$- and 2$N$-induced processes \cite{Gar03}.
It is overlaid on the top of Fig.\ \ref{fig:npene}
(dashed histogram) on the same scale.
The observed spectrum was significantly different from
the calculated energy spectrum.
The calculated spectrum still has its maximum at about Q/2,
whereas the experimental spectrum levels off
in the region from 20 to 80 MeV.

It should be noticed
that there are so far two types of the 2$N$-induced NMWD models.
One is a three-body reaction
in which the Q-value can be distributed over three nucleons
in the final state.
The other is based on the correlated two-nucleon absorption
of a virtual pion emitted from the weak vertex of $\it{\Lambda} N \pi$,
in which the energy of the emitted nucleon is small,
since the virtual pion is close to on-shell \cite{Gar03}.
Since most of the Q-value in this 2$N$-induced process
is distributed to two nucleons,
the shape of the energy spectra
should be close to that in the 1$N$-induced process
($\it{\Lambda} N \rightarrow n N$).
Thus, the latter model adopted in the calculated spectra
has a much smaller effect to enhance the low-energy region
than the former model.
This suggests the importance of the former 2$N$-induced model.
Another speculation could be that
the FSI effect is much stronger than
that in the calculation by Garbarino $\textit{et~al.}$

\section{Conclusion}

We measured the energy spectra
of neutrons and protons emitted from the non-mesonic decays of
$^{5}_{\it{\Lambda}}$He and $^{12}_{\it{\Lambda}}$C
with high statistics.
Both the neutron and proton spectra show a similar shape.
The obtained neutron-to-proton yield ratios, $N_n / N_p$,
for both hypernuclei
are approximately equal to two with high energy thresholds,
which are consistent with a simple estimation assuming
$\Gamma_n/\Gamma_p \sim$ 0.5.
This result rules out a theoretical calculation based on the OPE model,
and  supports recent calculations based on short-range interactions.
In the present experiment, the high statistics of the observed nucleons
has enabled us to carry out a coincidence analysis of the two nucleons,
$n+n$ and $n+p$, which is now in progress.
The analysis will provide a definitive $\Gamma_n/\Gamma_p$ ratio.
In the obtained neutron spectrum for $^{5}_{\it{\Lambda}}$He,
the spectral shape was not consistent
with the simple expectation
that there is a broad peak at one half of the Q-value.
This indicates the importance of the multi-nucleon induced process
in the NMWD or/and a large FSI effect, even for $^{5}_{\it{\Lambda}}$He.
This result calls for more detailed studies
concerning the 2$N$-induced NMWD process and the FSI effect.

\section*{Acknowledgments}
\label{acknowledgments}

We are grateful to Prof.\ K.\ Nakamura and the KEK-PS staff for support
of the present experiment.
We also acknowledge the staff members of the cryogenic group at KEK.
This work was supported in part
under the Korea-Japan collaborative research program of
KOSEF(R01-2000-000-00019-0) and KRF(2003-070-C00015).

\end{document}